\documentclass[acmsmall]{acmart}
\AtBeginDocument{%
  }

\usepackage{xcolor}
\usepackage{multirow}
\usepackage{xr}
\usepackage{graphicx}
\usepackage{subcaption} % for imaes

\newcommand{\edit}[1]{\textcolor{black}{#1}}

\usepackage{enumitem}

\begin{document}

\author{Alice Qian}
\email{aqzhang@andrew.cmu.edu}
\affiliation{%
  \institution{Carnegie Mellon University}
  \country{USA}
}

\author{Ryland Shaw}
\authornote{Ryland Shaw was a pre-doctoral research assistant at Microsoft Research (Cambridge MA) at the time of this writing.}
\affiliation{%
  \institution{University of Southern California}
  \city{Los Angeles}
  \state{California}
  \country{USA}
}

\author{Laura Dabbish}
\email{dabbish@andrew.cmu.edu}
\affiliation{%
  \institution{Carnegie Mellon University}
  \country{USA}}

% TO DO 
\author{Jina Suh}
 \authornote{Contributed equally as co-last authors.}
\email{jinsuh@microsoft.com}
\affiliation{%
  \institution{Microsoft Research}
  \country{USA}}

\author{Hong Shen}
\authornotemark[2]
\email{hongs@andrew.cmu.edu}
\affiliation{%
  \institution{Carnegie Mellon University}
  \country{USA}}



\renewcommand{\shortauthors}{Alice Qian, Ryland Shaw, Laura Dabbish, Jina Suh, and Hong Shen}

\begin{abstract}
As AI systems are increasingly tested and deployed in open-ended and high-stakes domains, crowdworkers are often tasked with responsible AI (RAI) content work. These tasks include labeling violent content, moderating disturbing text, or simulating harmful behavior for red teaming exercises to shape AI system behaviors. 
While prior research efforts have highlighted the risks to worker well-being associated with RAI content work, far less attention has been paid to how these risks are communicated to workers by task designers or individuals who design and post RAI tasks.
Existing transparency frameworks and guidelines, such as model cards, datasheets, and crowdworksheets, focus on documenting model information and dataset collection processes, but they overlook an important aspect of disclosing well-being risks to workers. 
In the absence of standard workflows or clear guidance, the consistent application of content warnings, consent flows, or other forms of well-being risk disclosure remains unclear.
This study investigates how task designers approach risk disclosure in crowdsourced RAI tasks. 
Drawing on interviews with 23 task designers across academic and industry sectors, we examine how well-being risk is recognized, interpreted, and communicated in practice. Our findings highlight the need to support task designers in identifying and communicating risks not only to support crowdworker well-being but also to strengthen the ethical integrity and technical efficacy of AI development pipelines. 

\end{abstract}

%  TO DO add ccs

% Articles V10cscw001-V10cscw043 use
\received{May 2025}
\received[revised]{November 2025}
\received[accepted]{December 2025}

\title[Locating Risk]{Locating Risk: Task Designers and the Challenge of Risk Disclosure in Crowdsourced RAI Content Work}

% \title{Navigating Risk and Responsibility in AI Task Design}
% title options: RISE (Risk-Informed, Safe, and Ethical AI Task Design), RAID (Risk Accountability in AI Design), Foundations for Safe AI Task Design, Navigating AI Risks: A Framework for Responsible Task Design, Responsible AI Task Design, Balancing Act: Safe and Responsible AI Task Design
\maketitle

\section{Introduction}

% SECTION 1: What is the problem? Why is it important?
A task titled ``Emotion Detection in Text Conversations'' appears on a popular crowdsourcing platform. The description advertising the task is brief: workers will read excerpts from online chats and label the emotional tone. A crowdworker clicks ``Start,'' expecting something benign. Instead, they are shown messages sent to a suicide crisis hotline. Some include graphic references to self-harm and despair. The worker finishes the task but later writes to the task designer, saying the experience triggered symptoms of their post-traumatic stress disorder (PTSD) and left them unable to sleep for days. They write, ``I wish there had been a warning. I never would have clicked on this if I had known.'' For the task designer, the task was part of an effort to build artificial intelligence (AI) models for mental health support. But for the worker, the lack of meaningful disclosure turned a paid task into an unexpected site of personal harm. This composite scenario illustrates well-documented risks reported in media accounts and academic research on crowdworker exposure to sensitive, distressing, and potentially harmful content~\cite{mann2025meta}. \edit{As \textbf{``responsible AI (RAI) content work''} \cite{zhang2024aura} becomes an increasingly important part of the infrastructure of AI development \cite{zhang2025effective, udupa2023ethical}, this scenario raises questions about who is responsible for the well-being of crowdworkers, and how \textbf{``well-being risks''} are weighed, mitigated, or disclosed, if at all ~\cite{bharucha2023content, pinchevski2023social}.}

% human labor underpins AI development but is often overlooked
%  SECTION 2: what's been tried/why it didn't work

\edit{The work of making AI ``responsible," as elsewhere in the AI development process, requires massive amounts of human experience and judgment. Much of this work is facilitated through crowdsourcing platforms~\cite{gray2019ghost, roberts2016commercial, zhang2024aura}.} Prior work in HCI and CSCW has advanced transparency as a key strategy for mitigating risks to workers in responsible AI development, particularly through documentation practices that clarify how systems and datasets are created and used. \edit{Efforts to advance transparency in AI development have largely focused on reporting macro-level labor decisions \cite{mitchell2019model, gebru2021datasheets, diaz2022crowdworksheets} or building tools to help workers protect their time and income \cite{irani2013turkopticon}. Although these efforts represent meaningful progress in improving conditions for RAI content workers, they do not directly address the well-being risks that workers may face while completing tasks, nor do they take into account how task designers specifically make situated, interpretive decisions about when and how to be transparent about the risks associated with a task. Disclosing such risks during RAI crowdwork tasks remains ad hoc, structurally unsupported, and under-researched. Even though some platforms, such as Prolific, offer tools for content warnings and guidance for disclosing sensitive content, there is no regulation or standardization in how similar tools are integrated across other platforms. Furthermore, it is unclear whether and how these tools are used in practice, if they meaningfully support worker well-being, and how they may affect the quality of data produced by workers. 
This ambiguity results in risk disclosure being viewed as a matter of individual judgment rather than a shared responsibility.}

We propose that a crucial step to take in addressing this gap is to better understand the role of the ``\textbf{task designer},'' the individuals who design and post RAI tasks. This role has been traditionally labeled as \textit{requesters} in the crowdsourcing literature, but such a term implies a transactional role centered on task submission and payment. In contrast, \textit{task designer} emphasizes the creative and interpretive decisions involved in shaping task content, instructions, and interaction flow ~\cite{zheng2011task, bragg2018sprout, alagarai2014cognitively, allen2018design}. 

This study examines how task designers posting tasks on open crowdsourcing platforms make decisions about whether, when, and how to disclose well-being risks to workers involved in RAI content tasks. Specifically, we ask:
\begin{itemize}
    \item \textbf{RQ1:} How do task designers consider and communicate well-being risks to crowd workers in RAI content tasks?
     \item \textbf{RQ2:} What contextual factors and tensions influence task designers’ sense of responsibility, and how do these factors and tensions shape their decisions about disclosing risk to crowdworkers?
\end{itemize}

In doing so, our study contributes a grounded understanding of how task designers navigate risk disclosure in crowdsourced AI development. We find that decisions about whether, when, and how to disclose well-being risks unfold across three key stages of task design: (1) conceptualization, (2) specification, and (3) evaluation, each marked by a lack of shared standards. Task designers relied on personal assumptions about harm thresholds, varied widely in how they communicated risks, and often inferred harm through indirect worker signals after the task was complete. These practices were further shaped by organizational norms, platform affordances, and broader professional cultures, which encouraged ethical gestures but rarely \edit{levied} accountability. Finally, many task designers perceived risk disclosure as a tradeoff with data quality, revealing key tensions between protecting worker well-being and meeting the demands of rigorous, scalable RAI development. 

Based on these findings, we argue that risk disclosure in crowd-based RAI content work is not only an ethical decision, but also a structurally dislocated one \cite{widder_dislocated_2023}---with responsibility diffused across individuals, platforms, and institutions without clear ownership or accountability. %This dislocation places a disproportionate burden on task designers, who are often left to navigate worker well-being risks without adequate guidance or support. 
We call for repositioning the task designer as a critical point of intervention for mitigating these risks. We also advocate reimagining risk disclosure as a design and infrastructural challenge, one that requires integrated tools, shared norms, and review practices that move beyond ad hoc, individually improvised approaches.

\section{Related Work}

\subsection{Well-being Risks in RAI Content Work}
As AI systems' reliance on human-generated data means that content workers are frequently tasked with reviewing, annotating, or generating content that involves hate speech, harassment, violence, or trauma-related material. These content workers, including annotators, moderators, and adversarial red teamers, face risks not just of cognitive fatigue but of great psychological and physical harm. Studies have documented a range of effects: secondary trauma and emotional burnout \cite{wohn2019volunteer, dosono2019moderation}, privacy violations \cite{pinchevski2023social, schopke-gonzalez_why_2022}, belief system shifts \cite{newton_trauma_2019, Stackpole_2022, Douek_2021}, and even PTSD \cite{Michel2018ExContentMS, ruckenstein_re-humanizing_2020, Dwoskin_2019, arsht_2018_human}.

While financial and structural risks have received sustained attention \cite{silberman2018responsible, gray2019ghost, xu2017incentivizing}, risks to workers’ well-being as a consequence of exposure to sensitive or disturbing task content remains underexplored. The crowd work landscape has evolved over time, introducing new risks around the well-being of annotators themselves. As more routine annotation tasks become automated, human workers are increasingly tasked with complex and sensitive content, such as moderating hate speech, labeling trauma-related data, or simulating harmful behavior for adversarial testing ~\cite{zhang2024aura, gillespie2024ai, raji2020closing, gray2019ghost}.

\subsection{Worker Risk Mitigation}
HCI and CSCW scholars, in particular, have studied and proposed several design-based interventions and organizational safeguards and policies to support content and crowdworkers. At the same time, these tools and policies are not being adopted within organizations. A wider set of scholarship has started to situate the problem of poor crowdwork labor conditions in organizational dynamics, such as in corporate pressures to release new AI products, or in lackluster academic institutional review board processes. 

\subsubsection{Interface-level Mitigation Strategies}
To reduce exposure-related harm, researchers have proposed a variety of interface-level techniques, such as greyscale rendering \cite{spence2006color}, image blurring \cite{lin2009capture}, blocking facial expressions to reduce amygdala activation ~\cite{costafreda2008predictors, fusar2009functional}, and audio muting to limit sensory overload ~\cite{holmes_key_2010}. Post-task decompression strategies include playing Tetris to disrupt visual flashbacks~\cite{holmes_can_2009}, mindfulness exercises \cite{karunakaran2019testing}, and expressive writing or journaling \cite{das_fast_2020}.
Organizational and clinical solutions often focus on building long-term psychological resilience. These include resilience training programs \cite{steiger2022effects}, evidence-based psychotherapy for trauma \cite{cusack2016psychological, watts2013meta, cook2022awe}, and structured employee assistance programs (EAPs) or workplace wellness tools~\cite{spence2023content, steiger_psychological_2021}. Although effective, these solutions are largely designed for corporate contexts or formal institutions, and are rarely available to crowd workers or one-off annotators.

Despite the availability of these tools, many interventions are implemented only after harm occurs, rather than preventing it. In open, crowdsourced research settings, it is rarely required to frame tasks as involving human subjects, even when the work clearly entails psychological or emotional risk~\cite{xia2022original}. This leads to situations where workers encounter distressing content without any prior disclosure, raising ethical concerns about informed consent and the right to opt out.
While post-task support and exposure-reduction tools remain important, our work highlights clear, upfront disclosure as a critical and underutilized form of mitigation. Content warnings, pre-task descriptions, and consent flows are simple yet powerful tools—but they are inconsistently applied, and designers often lack guidance about when or how to use them. We argue that meaningful disclosure is not just a courtesy but a form of ethical infrastructure--one that enables worker autonomy, reduces harm, and fosters more responsible AI development from the ground up.

\subsubsection{Platform-Level Mitigation Strategies}
Crowdsourcing platforms, like social media sites, online marketplaces, and so on, have long argued that they are not responsible for harm that stems from how users behave on their platform \cite{Gillespie_politics}. This argument is premised on a supposed neutrality of algorithms and digital systems \cite{burgess_affordances_2018}, which discursively and legally shields platforms from having to thoroughly comply with labor regulations \cite{Wood_2019Disembeddedness_of_labor}.  However, crowdwork platforms can shape what tasks can be created, how workers engage, and the quality of results through the design of the platform's features, frictions, and limitations \cite{jarrahi_platformic_2020}.

\begin{figure}[ht]
    \centering
    \begin{subfigure}[t]{0.45\textwidth}
        \centering
        \includegraphics[width=\textwidth, height=0.6\textheight, keepaspectratio]{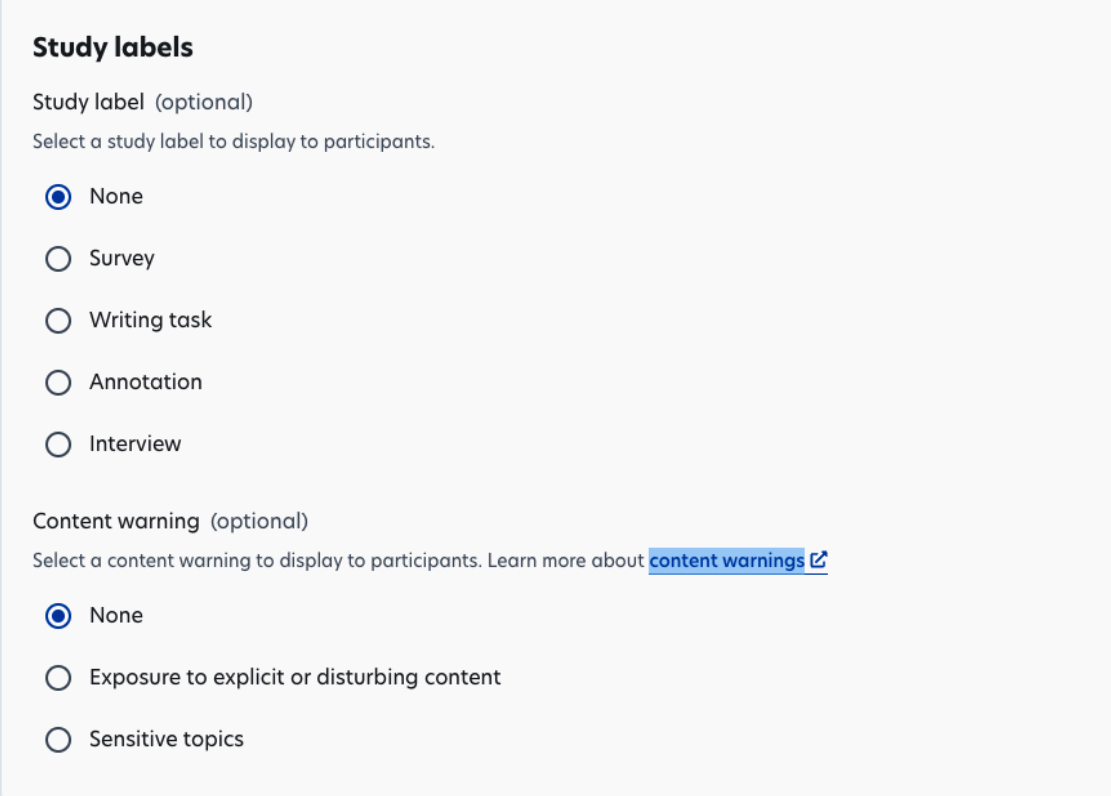}
        \caption{Selecting type of content warning}
        \label{fig:prolific-warning-type}
    \end{subfigure}
    \hfill
    \begin{subfigure}[t]{0.45\textwidth}
        \centering
        \includegraphics[width=\textwidth, height=0.6\textheight, keepaspectratio]{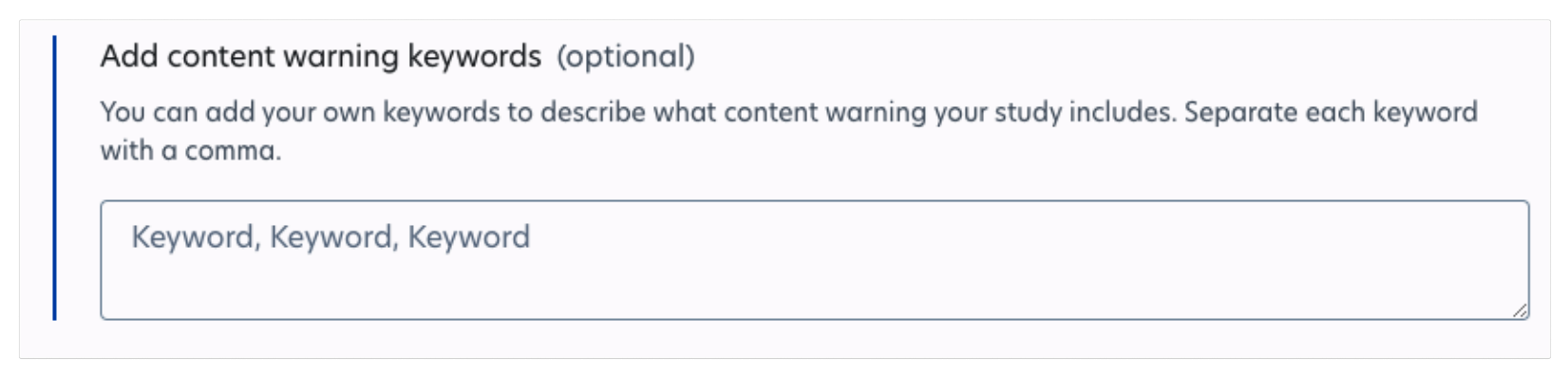}
        \caption{Adding content warning keywords}
        \label{fig:prolific-warning-keywords}
    \end{subfigure}
    \caption{Prolific's warning for sensitive content in tasks from the task designer interface.}
    \label{fig:prolific-warnings}
\end{figure}

Building on this view, prior HCI and CSCW work examines the structural dimensions of crowdsourcing platforms \cite{fieseler_unfairness_2019}. Foundational efforts, such as Turkopticon, have surfaced power asymmetries between workers and task designers \cite{irani2013turkopticon}. Subsequent studies have shown how tradeoffs between efficiency and fairness are embedded in workflows, incentives, and reputation mechanisms \cite{ho2015incentivizing, saito2019turkscanner}. Platform-level mitigation options remain limited and uneven, resulting in opaque rejection processes, emotional strain, and weak institutional protections \cite{mcinnis2016taking, flores2020challenges, martin2014being, salehi2018ink, silberman2018responsible}. In turn, the celebrated flexibility of crowd work frequently proves illusory \cite{rechkemmer2022understanding, varanasi2022feeling, liang2021embracing}. At the same time, data quality remains central for task designers, and choices about instructions, interfaces, and incentives co-produce both outcomes and worker experience \cite{wu2017confusing, han2020crowd, ho2015incentivizing}. These dynamics are structured by governance and algorithmic infrastructure at the platform level, which mediates access to tasks, enforces norms, and shapes who gets to work and under what conditions, often under claims of neutrality and scale \cite{toxtli2021quantifying, whiting2019fair, gray2016crowd, gadiraju2017modus, rzeszotarski2012crowdscape}.

A comparative view across platforms clarifies how different affordances distribute risk and agency. Prolific offers one of the more developed examples of platform-level support: during task creation, task designers are prompted to indicate whether a study involves explicit or disturbing content or sensitive topics, can supply keywords, and can consult explanation pages for further guidance \cite{prolific2025sensitive, prolific2025participant} (see Figure~\ref{fig:prolific-warnings}). Other systems restrict exposure through eligibility rules such as age-based limits or platform qualifications, as in the case of Amazon Mechanical Turk (MTurk) \cite{mturk2018aup}. Volunteer research platforms such as Zooniverse and Toloka document task-level warnings, though formats and enforcement vary across projects.\footnote{https://www.zooniverse.org/}\footnote{https://toloka.ai/} Some commercial annotation services, including DataAnnotation, Remotasks, and SurgeAI, further gate access behind employment or identity checks, and public descriptions of their disclosure practices remain limited. The comparative effectiveness of these approaches is unclear, as is how task designers perceive and enact them in practice.

Beyond warnings and qualifications, feedback and reputation systems mediate how task concerns travel between workers, task designers, and platforms. Prior work documents limited mechanisms for workers to pass feedback on tasks back to requesters and platform staff \cite{gegenhuber_microphones_2021, irani2013turkopticon}. At the same time, platforms rate workers through quality metrics, a form of algorithmic management \cite{lee_working_2015} that conditions access to tasks and pay, while equivalent in platform rating systems for task designers remain absent on MTurk and are scarce elsewhere \cite{fieseler_unfairness_2019}. Given the stakes of these ratings, workers may invest extra labor to protect scores and sometimes game reputation systems \cite{bellesia_algorithms_2023}. These governance choices intersect with content warning tools and qualifications, since both influence who sees which tasks, how concerns are escalated, and where responsibility for risk disclosure is perceived to reside.

\subsubsection{Organizational Ethics and Guidelines} Moving beyond transparency, there has been a proliferation of ethical guidelines intended to address the broader societal impacts of its development and deployment \cite{attard-frost_ethics_2023}. Google's AI Principles, for example, claim to uphold "responsible development and deployment" that follows "widely accepted principles of international law and human rights." \footnote{https://ai.google/responsibility/principles/} But because of the industry's legal norm of secrecy, it is difficult for researchers to examine how transparency initiatives and ethics principles are translated and operationalized for lower-level employees at tech companies like task designers, who work at the critical boundary between the tech company and its contracted crowdworkers. In one of the few research projects that examined this process, Ali et al. \cite{ali_walking_2023} found that tech employees struggled to meaningfully implement ethics guidelines as they grappled with the competing pressure to meet product deadlines, as well as with organizational structures and incentives that were largely incompatible with the thoughtful consideration of ethics and risks. 

\subsubsection{Academic Governance Mechanisms} Academic researchers and, less frequently, industry researchers,\footnote{Some tech companies have accredited IRBs, like Microsoft Research. Industry researchers may also contract with an external IRB, or partner with researchers at an academic institution to use their IRB. However, this pertains only to "research," which often needs to be approved by IRBs in order to be published in academic venues. Engagement with crowdworkers that is not intended for academic publication would likely go through more opaque company ``compliance'' checks, or would not be subject to any review at all.} must also pass rigorous checks by institutional review boards (IRBs), which seek to ensure that any research that interfaces with humans preserves their safety and privacy. However, Zook et al. \cite{zook_ten_2017} caution that, because so much computer science research happens outside governance mandates of IRBs and outside the expertise held by its assessors, researchers should hold their work to higher standards than IRBs request. This is easier said than done, given that there is little consensus among researchers about how ethical social computing research should be conducted \cite{vitak_beyond_2016}. Given that so much RAI crowdsourcing happens within corporate walls and well outside the purview of IRBs, many task designers are not likely to receive guidance from IRBs at all. 

\subsubsection{Transparency as a Mitigation Strategy}
While transparency tools like model cards~\cite{mitchell2019model}, datasheets for datasets~\cite{gebru2021datasheets}, and data nutrition labels~\cite{holland2020dataset} were originally developed to surface issues in model performance, data provenance, and intended use, these frameworks have also been positioned as mechanisms to mitigate RAI risks across the data pipeline. However, the risks they primarily address tend to focus on downstream users, developers, or consumers, leaving gaps in how transparency might protect the workers generating the training data. More recent tools like CrowdWorkSheets~\cite{diaz2022crowdworksheets} begin to bridge this gap by encouraging documentation of annotator diversity and task parameters. Yet even these frameworks treat disclosure as a reflective design consideration rather than an ethical obligation tied to worker well-being. Moreover, their adoption remains limited across academic and industry contexts, in part because they are not embedded in formal accountability infrastructures~\cite{turri2024transparency}.

 % why is it important and unique we focus on task designers in the context of wellbeing risk? 
\subsection{Task Designers and the Situated Nature of Risk Disclosure}

Task designers occupy a pivotal yet often ambiguous role within this ecosystem. They are responsible for shaping not only task content but also how risks are framed, communicated, or omitted. Task designers exercise discretion over whether to include content warnings~\cite{mason2012conducting}, how tasks are contextualized~\cite{finnerty2013keep}, and how tradeoffs between speed, cost, and ethics are negotiated~\cite{gaikwad2016boomerang, papoutsaki2015crowdsourcing, kittur2008crowdsourcing, xia2020privacy}. In the emergence of increasing demand for crowdsourced RAI content work, we surface a need to better understand task designers’ approaches to mitigating well-being risks for workers. Prior research in crowdsourcing and HCI has focused largely on worker experiences~\cite{irani2013turkopticon, salehi2015we, silberman2018responsible, williams2019perpetual} or platform infrastructure~\cite{jarrahi_platformic_2020}, leaving a gap in our understanding of the interpretive and discretionary practices of task designers themselves.

These decisions may be made without formal training or review processes, especially for first-time designers or those in smaller organizations~\cite{papoutsaki2015crowdsourcing}. Many rely on intuition, peer templates, or trial and error, leading to highly variable practices~\cite{gutheim2012fantasktic, finnerty2013keep, kittur2008crowdsourcing, papoutsaki2015crowdsourcing}. The ambiguity surrounding how and when to disclose risk underscores the need to examine how well-being concerns are recognized, interpreted, and acted upon in real-world settings. 

By studying RAI task designers as central actors and their practices, we surface the nuances of how well-being risk is recognized, negotiated, and communicated in practice from RAI task designers' perspectives. Our work contributes to the growing body of work on developing methods to support a robust human infrastructure for AI development~\cite{zhang2024aura, wang2022whose, gray2019ghost, irani2016stories} by shifting analytical attention to the mid-level task designers of crowdsourced tasks. 

\section{Methods}
To understand how task designers navigate risk disclosure in the context of crowdsourced RAI tasks, we conducted a qualitative interview study with 23 task designers. Semi-structured interviews are well-suited for exploring under-examined practices, enabling participants to reflect on their own decisions, constraints, and reasoning in their own terms~\cite{holstein2003inside, yin2015qualitative, smith1995semi}. This approach allowed us to assess the situated, often discretionary nature of risk-related choices, unlikely to surface through log-analysis or survey-based methods alone.

\subsection{Recruitment}
We recruited a total of 23 participants through social media postings on LinkedIn, publications, and referrals. To ensure participants had experience designing crowdsourced RAI content work tasks, we asked potential participants to provide information about their background (i.e., job title and sector) and the type of RAI tasks they previously designed. We provided predefined options based on prior literature on types of RAI content work~\cite{zhang2024aura} and types of sensitive content~\cite{zheng2011task, shevlane2023model, weidinger2023sociotechnical}.

We asked for the type of organization each participant belonged to (e.g., academia, industry), the type of tasks they typically request (e.g., data generation, content removal, adversarial prompting, etc.), and the type of sensitive content within their tasks (e.g., misinformation, violent and graphic content, etc.), and their location country. We also asked participants to provide optional examples of tasks they requested to further understand their experience. Each participant received \$30\footnote{One participant located in Portugal was paid the equivalent of \$30.} for their participation. The lead institution's institutional review board (IRB) approved the study protocol. 

In our recruitment and interview, we referred to the content shown within RAI content tasks as \textit{sensitive content}, although prior literature may use alternative terms such as potentially ``harmful'' or ``distressing'' ~\cite{bharucha2023content, zhang2024aura, solaiman_evaluating_2023}. We chose the term ``sensitive'' because it aligns with the language commonly used on crowdsourcing platforms like Prolific for content that may \textit{potentially} cause harm through exposure and is likely what task designers may be most familiar with~\cite{prolific2025participant, prolific2025sensitive}. To help participants determine whether their work involved sensitive content, we included concrete examples drawn from prior literature: data generation (e.g., submitting toxic language), data labeling (e.g., identifying sexual content), content removal (e.g., deciding whether content should be model-suppressed), and adversarial prompting (e.g., writing prompts likely to elicit violent model outputs).

While job titles were self-reported, we inferred participants’ organizational role levels (i.e., senior, mid-level, or junior) during analysis based on how they self-described their responsibilities, autonomy, and influence. Following prior work on organizational roles and professional authority~\cite{barley2001bringing}, we applied a broad schema: senior-level roles involved strategic influence or supervisory duties; mid-level roles reflected specialized skills with limited decision-making power; and junior roles included entry-level or trainee positions. This classification helped contextualize how participants' organizational position shaped their approach to risk disclosure (see Table~\ref{tab:participant_summary}).

\begin{table}[h]
\centering
\begin{tabular}{p{4cm} p{10cm}}
\toprule
\textbf{Category} & \textbf{Summary} \\
\midrule
Sector & Industry (16), Academia (7) \\
Job Titles& Professor (1), PhD Student (6), AI Researcher (4), Research Scientist  (1), Data Analyst (6), Data Scientist (3), Project/Product Manager (2)  \\
Organizational Role Level & Senior-level (7),  Mid-level (9),  Junior-level (7)\\
Types of Sensitive Content & Hate speech (17), Bias/stereotyping (16), Harassment and bullying (11), Violent and graphic content (9), Misinformation (4), Terrorism and extremism (3),  Scams and fraud (1)\\
Task Type & Data annotation (21), Content removal (14), Data generation (10), Adversarial prompting  (9), Content removal (1)\\
Platform Used & Prolific (19), Amazon Mechanical Turk (17), Remotasks (5), TaskUs (5), UHRS (2), Upwork (1), Clickworker (1) \\
Location & United States (22), Portugal (1)\\
\bottomrule
\end{tabular}
\caption{Summary of participant characteristics across sector, job titles, task types, sensitive content handled, and platforms used.}
\label{tab:participant_summary}
\end{table}

\subsection{Interview Procedure}

We conducted semi-structured interviews using a protocol designed to surface how task designers perceive and enact responsibility for worker well-being during crowdsourced RAI content work. Interviews were conducted remotely via video conferencing software and lasted between 60 and 90 minutes. Each session began with a review of the consent form, followed by verbal confirmation of consent. Participants were informed that they could skip any question or end the interview at any point. All interviews were audio-recorded with permission and then transcribed by Otter.ai.

The interview protocol was organized into five main sections: (1) background and context, (2) task setup and motivations, (3) task information and communication, (4) post-launch task management, and (5) debriefing and challenges. In the first section, we asked participants to describe their professional background, define Responsible AI in their own terms, and explain how they became involved in RAI content work. These opening questions helped situate each participant’s perspective on AI and worker responsibility.
The core of the interview focused on walking through a specific example of a task the participant had designed that contained sensitive content. Participants were prompted to describe the purpose of the task, how it was set up, what platform it was deployed on, and what information was included in the task description. \edit{Participants were asked to optionally screen-share examples of tasks documented (e.g., in their publications) to better enrich discussions.} Follow-up questions explored how participants thought about worker experience and risk during the design phase. We probed decisions about what to disclose to workers, how warnings (if any) were written, and whether institutional or platform-level resources were consulted.
Later sections asked participants to reflect on what happened after a task was posted: how they monitored worker responses, whether they received any complaints or feedback, and if or how they adjusted the task in response. Participants were also asked about their use of communication channels with workers, including whether they allowed message-based feedback or requests for task rejection appeals. In the final portion of the interview, participants reflected on broader challenges they faced in disclosing risk, managing transparency, and balancing competing priorities such as data quality, ethical obligations, and platform constraints.

Our protocol was inspired by prior work on content moderation~\cite{roberts2019behind}, Responsible AI practices~\cite{raji2020closing}, and documentation frameworks~\cite{diaz2022crowdworksheets}, and iteratively refined through pilot testing with colleagues. The open-ended structure allowed us to ask follow-up questions to obtain more depth on some topics while maintaining consistency across interviews. The goal was not only to document disclosure practices but to understand the rationales, tensions, and constraints that shaped them.

\subsection{Data Analysis}
We anonymized and coded all 23 transcripts using thematic analysis~\cite{clarke2017thematic, smith1995semi}. First, we created precise, low-level codes for each transcript, then we organized these codes into axial codes. Afterward, we iteratively coded transcripts through the steps of consolidating codes, discussing them, identifying themes, and writing memos. Initial codes were created and reviewed by the first two authors, while the remaining authors were consulted during the iterative refinement of themes until saturation was reached in the data, at which point new interviews no longer provided new insights. 

Our analytical focus was not only on what designers disclosed, but also on how they framed those decisions: what they considered to be risky, what shaped their disclosure strategies, and how they understood their role in relation to worker well-being. This reflexive, interpretive approach allowed us to attend to the situated reasoning of task designers and the sociotechnical environments in which they operate.

\section{Findings}

%  show through subheads lack of standards etc.
%  new organization
%  (1) what - what is the risk
%  (2) how - how are decisions made and risk is communicated
%  (3) why - tradeoffs and considerations

\begin{figure}
    
    \includegraphics[width=\textwidth, height=0.5\textheight, keepaspectratio]{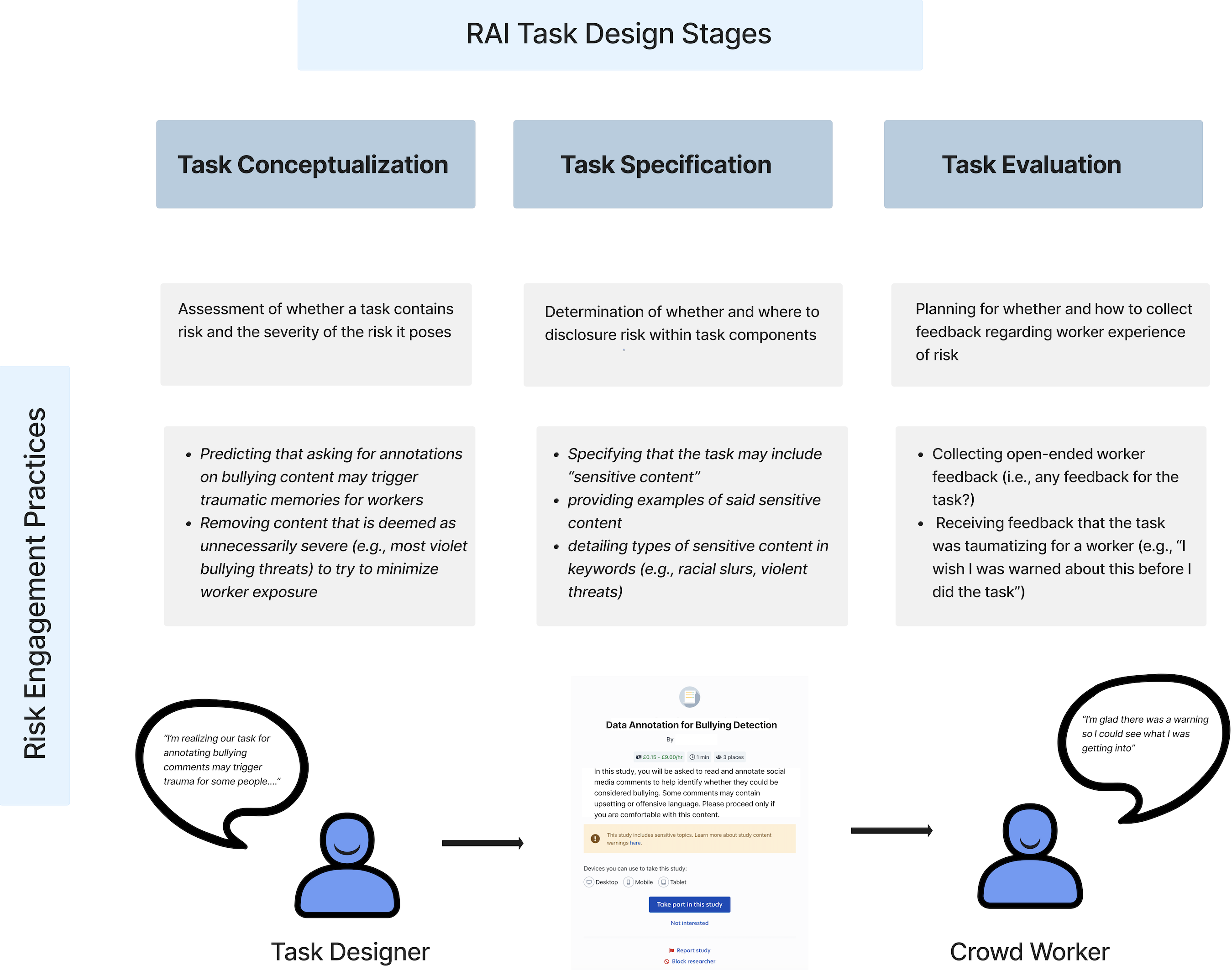}
   
    \caption{Risk-related decision-making across stages of AI task development.}
    \label{fig:risk-matrix}
\end{figure}

In this section, we examine how \edit{our task designer participants}  navigated the communication of well-being risks when designing and deploying crowdsourced RAI tasks. Task designers often felt responsible for protecting workers but lacked standardized tools or norms to guide when, where, and how to disclose risk. As a result, task designers described making case-by-case decisions shaped by personal ethics, perceptions of risk, platform affordances, and professional norms. Figure \ref{fig:risk-matrix} summarizes how each of these dimensions plays out across stages. 
%\edit{We note our analysis illustrates practices and challenges faced by our particular selection of participants, who, due to the nature of the study, may be more likely to have reflected on challenges faced by workers prior to participation in the study.} 
% \ryland{can someone clarify what "more likely to reflect on challenges faced by workers" means? Is this to say our selection might have biased the group toward people who are more concerned or more reflexive about their relationship to crowdworkers than is typical?}

\subsection{Risk Disclosure Across RAI Task Design Stages}

Overall, we found that task designers made interpretive, context-sensitive decisions about whether, when, and how to disclose well-being risks across the stages of AI task development, using few standardized norms or shared definitions to guide these decisions. \edit{To provide clarity in the organization of our findings, we bring forth an organizational framework of three AI task design stages that have been discussed in prior research on crowdwork ~\cite{ho2015incentivizing, zheng2011task} and content moderation ~\cite{bharucha2023content}, but have received limited discussion in an AI context.} \textbf{Task Conceptualization} involves the initial deliberation on the overall idea, scope, and framing of a task. \textbf{Task Specification} refers to the stage at which the components of tasks are written out with specifics such as the pre-screening and consent information, the title of the task, and the task description. \textbf{Task Evaluation} refers to the evaluation of the success of an AI task, including the analysis of worker input in the task and worker feedback about the task. 
% These AI task design stages emerged from our data, but aligned with the recruitment, task execution, and evaluation stages of tasks presented to content workers in non-crowdwork, employed settings~\cite{bharucha2023content}. 
We observed that within the RAI crowdwork environment, these three stages occur at a task-level: rather than eliciting recruitment for a group of workers for a multitude of tasks, task designers recruited workers for one specific task. Components in task specifications were developed specific to each task as well as the evaluation measures. 

\subsubsection{Framing Tasks Around Perceived Worker Thresholds} 
During the Task Conceptualization stage, task designers made two interrelated but distinct types of assumptions: (1) assumptions about the psychological or emotional impact of task content, and (2) assumptions about the thresholds of harm that workers could tolerate. These assumptions shaped how designers described tasks, whether they included content warnings, and how they conceptualized risk from the outset.

First, task designers made subjective judgments about the likely impact of task content on workers. At a baseline, most acknowledged that even brief exposure to sensitive material could be emotionally taxing. Some predicted specific emotional or psychological consequences. For example, P21 anticipated that asking workers to annotate bullying comments to train an AI model might trigger traumatic memories, while P19 noted the risk of frustration or burnout from repeated exposure to emotionally charged material.

Second, task designers drew on normative assumptions about worker tolerance, meaning the kinds or amounts of distressing content they believed workers could reasonably endure without harm. This notion of a `tolerance threshold' varied \edit{among task designers depending on the strategies they employed to make the judgment.} P1, for instance, assumed that most crowdworkers would be desensitized to harmful content due to regular internet use, stating: \textit{``[crowd workers are] not getting new revelations about how the internet works.''} In contrast, P8 adopted a more cautious stance, assuming workers might have a much lower threshold for harm. P8 described reviewing and calibrating content with their research team to \textit{``determine the exact type of content threshold [of harm] that [they] wanted to use for different pieces of content.''} Others fell in between: P12, for example, assessed whether the proportion of negative content in a task was high enough to merit a warning, signaling a heuristic approach to estimating harm tolerance. Together, these two concepts of impact and tolerance shaped how risk was described and justified at the earliest stage of task design. Yet both were informed by task designers’ own intuitions, rather than shared standards or direct input from workers themselves.

\subsubsection{\edit{Distinguishing Implicit, Vague, and Explicit Warnings of Task Risk}}

% Disclosure as warnings in description
Task descriptions emerged as the primary site for communicating risk, but task designers \edit{held differing understandings and enactments of that responsibility, which were based on platform features available to them as well as individual professional judgements.} \edit{For some, risk disclosure meant providing an accurate task description itself. Others believed best practices warranted including vague cautionary language while a few task designers strictly felt the need for }formal content warnings that explicitly detailed specific risks. These differences reflected not only stylistic variation in how task designers we spoke to specified components of their tasks, but deeper assumptions about what counts as meaningful risk disclosure and where the boundary then lies between providing a description of a task and including a warning. 

%  implicit disclosure 
Some designers believed that a clear and honest task description was sufficient to fulfill their risk disclosure obligations. \edit{P17 expressed that as a data analyst}, \textit{``I don’t think it’s actually a warning because this is just me telling [the workers] what the task is all about,''} suggesting that naming the task’s content was enough to signal potential harm. Others opted to signal risk through metadata rather than descriptions. For example, P22 listed keywords like ``toxicity'' and ``minority groups'' to hint at the nature of the content \edit{because the platform they used, MTurk, provided a field for keywords.}
%  vague warnings
Other designers included language meant to suggest the presence of sensitive material without naming it outright. P3 described this as offering \textit{``a content warning [that] there might be sensitive or controversial material,''} treating the phrase as a general signal without specifying the actual content. These disclosures left interpretation up to the worker and were often influenced by concerns about discouraging participation or biasing task engagement.

%  formal warnings that are EXPLICIT
In contrast, some designers treated the task description or instructions as a space for formal, explicit warnings. P22 clearly stated that the task involved hate speech or stereotypes and explained why the material might be distressing. Others focused not on content categories but on potential psychological outcomes. P16, for example, warned that the task could provoke emotional responses such as ``fear and depression.'' These designers framed warnings as ethical signals to support informed consent and reduce harm, often placing them alongside or within instruction pages.

% Disclosure as consent mechanism
For many task designers, consent forms and opt-out prompts served as key mechanisms for risk disclosure—offering workers the ability to make informed choices about participation before or during exposure to sensitive content. These moments were framed as ethical checkpoints rather than mere procedural steps, giving workers more control over their engagement. For example, P7 described a pre-task consent process in which workers were asked explicitly whether they wished to proceed: \textit{``If you answer yes, you move forward to the task itself. If you answer no, you'll be disqualified from this particular study, and can choose to withdraw without any penalty.''} Similarly, P5 described a mid-task opt-out option: \textit{``We are at this stage [of the task]. If you want to still continue, or you want to step off... the study will end there [if not], and you may continue.''}

\edit{Rather than use existing platform features,} other designers built \edit{customized} risk disclosures into the task interface itself, embedding cues such as example content, opt-out reminders, or reassurance that workers could exit at any time. These mid-task signals were intended to minimize harm in real time. For instance, P22 stated that they would \textit{``specify to workers that if they quit the task in the middle, it doesn’t affect them negatively... they can leave at any time.''} However, they also clarified that early exit would mean forfeiting compensation, revealing ethical tensions between task exit flexibility and task completion incentives. Not all task designers implemented such support. Many did not mention structured mechanisms for checking in on worker well-being during the task or offering ways to exit gracefully. Ultimately, these consent and workflow-based strategies reflect a broader orientation toward disclosure as an ongoing ethical process that extends beyond the start of the task \edit{influenced by task designers' individual judgments as well as platform affordances.}

\subsubsection{Evaluating Harm Through Ad-hoc Feedback or Inference}
After tasks were launched, designers either inferred worker harm through their task behaviors or collected worker feedback selectively, with only a few offering structured support or follow-up during the \textbf{Task Evaluation} stage. The common approaches for evaluating risks corresponded with known approaches for evaluating the success of the task overall. Primarily, task designers evaluated risk indirectly by observing behavioral patterns or response data to infer whether workers had been negatively affected, or took a more direct approach by collecting feedback from workers to inform future task design. A few task designers also engaged with post-task debriefings to provide workers with resources for support.

%  insert section on automated methods
Task designers commonly analyzed data from workers to \edit{determine if issues with the quality of worker input, such as the level of detail in written responses, could indicate that a worker was experiencing negative reactions due to} the content. 
Beyond formal feedback, a few task designers also engaged in passive evaluation, monitoring completion rates, response patterns, or unusual behavior to gauge whether something in the task might have been upsetting. P13, for instance, found that when they failed to adequately warn workers, they observed a higher volume of input that was short and vague, potentially indicating workers chose to be less attentive with the task. 

While several task designers intentionally sought and actively reviewed open-ended feedback from workers, others \edit{refrained from doing so because they felt the feedback would not be useful or applicable to future tasks. As such, P12} felt that feedback would not be applicable to future tasks if the task was designed to focus on a specific product (P12). Others believed that workers were more likely to use feedback forms to express dissatisfaction rather than constructive input. P10, for instance, stopped collecting feedback from workers because they believed that workers would exaggerate complaints about the difficulties they had in completing a task, simply because the option to provide feedback was given. \edit{Notably, P10 reached this conclusion based on their personal professional judgment, rather than consulting with team members. }

However, for those who did engage with feedback, both positive and negative responses were seen as highly informative. One participant described how a worker’s comment shifted their awareness of the emotional impact of their task: \textit{``[In] past experiences a couple of [workers] told me they didn't actually expect [the task] to be like this...it's giving them reminiscent images in [their] head. One time someone told me that my content gave him PTSD, and I was like wow.''} (P6)
Others saw feedback as a way to validate their decisions around disclosure. P15, for example, described how a worker responded positively to a pre-task warning: \textit{``[They] were glad [the warning] was mentioned...the fact that they can decide if they will be willing to go ahead with the task or not.''} \edit{Overall, task designers who sought feedback described it as helpful for shaping future disclosure decisions, while others chose not to solicit feedback because they doubted its value.}

A small number of task designers described offering post-task resources or debriefings as a way to address lingering emotional impact and to reinforce the value of the worker's contribution. This reflected a more sustained and intentional approach to supporting worker well-being after task completion. For example, P16 described offering workers access to support services as part of the study protocol: \textit{``We provide, you know, [a] helpline...where workers can actually reach out in case the sensitivity really affects them...some online resources. They can, reach out for help.''} In this case, post-task care was not only a reactive measure but also a recognition of the potential emotional toll of the work, and a way to extend ethical responsibility beyond task boundaries.

Together, these practices reflect that post-task engagement with risk is not uniform due to differences in opinions about its usefulness: some designers proactively sought out signals of harm, believing it could help them improve future tasks, while others avoided it altogether, questioning its usefulness or relevance. The variability in these approaches illustrates broader uncertainty around when and how responsibility for worker well-being extends beyond task completion.

\subsection{Risk Disclosure as a Negotiated Practice}
\begin{table}[h]
\centering
\begin{tabular}{|p{3.5cm}|p{10cm}|}
\hline
\textbf{Factors} & \textbf{Influence on Risk Disclosure} \\
\hline
Individual & Personal ethics and prior experiences shape whether task designers anticipate harm and decide to include risk disclosures.  \\
\hline
Organizational & Internal policies, such as IRB processes or company guidelines, provide ethical benchmarks. However, inconsistent enforcement can make risk disclosure feel optional or unsupported. \\
\hline
Platform & Tools like content warnings, participant filters, and messaging channels influence how and when designers communicate risk. Platform culture and norms also affect perceptions of ethical responsibility. \\
\hline
Community Culture & Community expectations around research ethics, publication standards, or disciplinary norms can drive decisions to disclose, especially in academic contexts. \\
\hline
\end{tabular}
\caption{Contextual factors shaping risk disclosure practices.}
\label{tab:contextual_factors}
\end{table}

While task designers exercised great personal agency in shaping how risk is communicated, their decisions were often entangled with the expectations, incentives, and constraints imposed by these surrounding systems. Task designers’ decisions about whether and how to disclose risk were shaped not only by personal values, but also by a layered set of influences from their organizations, platforms, and broader professional norms, making disclosure a negotiated and context-dependent practice.

\subsubsection{Interpreting Risk Through Personal Ethics}
Individual task designers relied heavily on their personal ethics and prior experiences to make decisions about disclosure, leading to highly varied interpretations of when risk disclosure was necessary or justified.
For some task designers, ethical values were a primary driver. P15, from industry, stood strongly on the topic of ethics, saying that \textit{``it's actually a big deal for me, if [a fellow task designer doesn't] care about the ethics...I feel [crowdsourcing] should be done ethically.''} P5, an academic researcher, similarly stated that they felt it was \textit{``more ethical to always provide clarity,''} but also added that they felt providing workers with more clarity about potentially sensitive tasks could \textit{``help the workers be more interest[ed] in doing [the task] rather than just doing it [by guessing].''} Some designers proactively put themselves in the shoes of the crowdworker to imagine how sensitive tasks might be received (P5, P16). P16 argued that every designer should have to \textit{``experience [the task] yourself''} before putting it online.  In these examples, an attitude of ``entrepreneurial ethics'' \cite{ali_walking_2023} served as a strong internal motivator.

Some task designers described becoming more attentive to well-being risks only after receiving negative feedback from workers---a form of reactive ethical concern. P7 recalled emails from workers who felt unprepared and asked for content warnings in future tasks: \textit{``I remember some emails I received...something like, `I wasn't really prepared for the type of sensitive context in this task,’ and they also said a content warning would have been helpful upfront.''} Others framed the inclusion of warnings less as ethical commitments and more as preemptive measures to avoid complaints. As P21 put it, \textit{``We do also put the warning just in case somebody complains in the feedback sections.''} These responses show how task designers’ ethical positions were shaped not only by internal values, but also by how they interpreted worker reactions, and whether they considered those reactions legitimate. These accounts highlight how, in the absence of formalized standards, task designers relied on personal ethics and worker feedback to shape their disclosure practices. However, this reliance on individual judgment contributed to inconsistent and reactive risk communication strategies.

%  inconsistent practices adaptation
% \subsubsection{Between Organizational Expectations and Ethical Accountability}
\subsubsection{Navigating Inconsistent Organizational Accountability}

Organizational guidance around risk disclosure ranged from robust internal standards to informal norms, with enforcement often lacking, leaving designers to interpret and implement expectations with wide discretion. In industry settings, some designers described formal company policies that included guidelines around informed consent and risk disclosure. 

In industry, P19 described how they had specific company guidelines to follow: \textit{``the [company] guidelines cover informed concepts, ensuring that the workers are fully informed about the nature of the task, potential risk and benefit,''} also noting in turn that their company \textit{``guidelines are informed by industry best practices, regulatory requirements, and ethical standards.''}  In one stand-out example, P16 noted that their company has a quality assurance department that reviews every part of the crowdwork task before it is deployed, although this type of internal check appeared to be unique among corporate task designers. However, the presence of guidelines did not always guarantee enforcement. P13 noted that while similar policies existed in their organization, ``there would be no consequences'' for failing to inform workers, though they believed doing so would reduce the quality of the task. Few other task designers mentioned formal accountability mechanisms if these guidelines were not met. 

Academic task designers, by contrast, often connected their organizational expectations to ethics review processes and academic publications' ethics compliance standards. Getting IRB approval required an important `check' at the beginning of a project, but did not contribute to continual or post-research oversight. Without strong enforcement mechanisms, organizational policies served more as soft signals than binding expectations. This ambiguity left task designers to navigate risk disclosure with discretion, resulting in wide variability across contexts.

\subsubsection{Reshaping Platform Affordances within Design Workflow}

Platforms shaped disclosure not only through technical affordances like filters and warning tools, but also through their perceived ethical reputation, which influenced how seriously designers treated their responsibilities. At the same time, task designers pointed to affordances, like pre-screeners, partial compensation tools, embedded warning displays, and feedback mechanisms as factors encouraging better risk communication. P6 noted that, unlike Mechanical Turk, Prolific made it easy to add content warnings, while P22 appreciated Prolific’s interface for compensating excluded task designers more fairly. Screening tools also helped task designers select populations of workers they felt would be most appropriate to complete certain sensitive tasks.

Yet not all affordances were used uniformly. P18 chose not to use pre-screeners to ensure inclusion of marginalized voices, showing how even ethically motivated decisions might bypass platform tools. Others noted the limits of algorithmic transparency on platforms like MTurk, where algorithmic ranking and marketplace dynamics could discourage disclosure. As P8 described,\textit{``there's a not-so-crazy world where if the marketplace doesn't like your survey, it can downrank your survey...without [you] knowing exactly what [the reason] is.''}

Moreover, some task designers found these features to be insufficient. P15 pointed out that the platform they used offered a feedback form at the end of the task, but that workers who exited a task early---possibly because they found the task too sensitive---were not offered this feedback form. Also recognizing this problem, P16 repurposed one of the platform's survey tools to force a feedback pop-up if a worker tried to exit the task early. Here, we see an ad-hoc, ``off-label'' \cite{duguay_you_2020} effort to develop better risk communication beyond the intended capabilities of the platform's affordances. 

These examples suggest that while platform features matter and can encourage or discourage certain ethical behaviors, ultimately task designers exercise their agency to make the platform's affordances \textit{work for them}. Even when tools for disclosure are built into a platform, how or whether they are used depends on the designer’s priorities, assumptions about risk, and the influences of the organization they are situated within. Platform affordances shaped risk disclosure practices, but task designers ultimately reinterpreted and adapted these tools based on their goals and constraints. This improvisation points to a need to design platform features that are both usable and contextually meaningful to task designers.

\subsubsection{Performative Risk Disclosure in a Culture of Urgency}
Task designers operated not just within organizations or platforms, but within the larger professional world of AI, a field defined by its rapid pace of development \cite{ali_walking_2023} and increasingly high standards for data quality and precision \cite{zhang2025effective, Sambasivan2021}. In this broader context, designers encountered community norms that encouraged risk disclosure rhetorically but rarely supported it structurally. Whether in academia or industry, designers felt pressure to act ethically, but they had to do so without shared standards or enforcement mechanisms. Disclosure thus became a flexible, often symbolic gesture, unevenly practiced depending on the task designer's professional community.

Across both academic and industrial settings, designers highlighted the pace and culture of AI development as a significant obstacle to sustained ethical reflection. Responsible risk disclosure, though valued, was often viewed as a potential friction in workflows optimized for speed. Several task designers described how filters to screen out potentially vulnerable workers, such as sensitivity flags or opt-out mechanisms, could delay data collection or reduce completion rates. P18 and P12 explicitly noted that extensive screening or disclosure steps risked shrinking the worker pool and jeopardizing progress for their overall AI development projects. This reflects a well-documented tension in responsible AI work: that efforts to design with care are frequently undermined by broader demands for rapid development~\cite{ali_walking_2023}. Ultimately, the speed-oriented culture of AI development often positioned risk disclosure as a hindrance to task designers, making it a symbolic rather than substantive practice. As a result, task designers were left to reconcile pressures for rapid delivery with the moral imperative to mitigate harm.

\subsubsection{Professional Ethical Standards Shaping Risk Awareness}
While urgency and speed often discouraged in-depth engagement risk, task designers' awareness of disclosure was also shaped by professional ethical norms such as those tied to institutional review processes, publication expectations, and informal professional networks. These norms often created pressure for task designers to appear to mitigate worker well-being broadly, but rarely offered meaningful support or enforcement.

In academic contexts, task designers described how peer review, IRB approval, and disciplinary ethics cultures shaped their sense of responsibility. However, these influences often encouraged ethical signaling rather than mandating specific actions. IRB processes were described as one-time hurdles rather than ongoing supports. All task designers in academia furthermore shared that their research was deemed as ``exempt'' from stricter rules regarding human subject research by their IRBs, despite the task designers being acutely aware that they were working with people. Exemption from human subjects research rules is common for crowdwork-based research projects in the field of machine learning, but scholars have recently contested this norm \cite{kaushik2023resolvinghumansubjectsstatus}. While task designers didn't need to follow the stricter rules governing human subjects research, at the minimum their awareness of the process appeared to make them more mindful of risk. 

Within professional communities such as journals or conferences, risk disclosure was sometimes motivated less by institutional accountability and more by the desire to avoid friction during the peer review process: \textit{``there’s a disincentive...like a carrot and stick. Your paper won’t appear at the conference''} (P8). While these academic norms helped raise awareness about risk, they did little to support designers in making context-sensitive decisions about disclosure. In some cases, task designers described including vague or general warnings simply to preempt reviewer concerns (P22).

% Industry Norms: Informal Learning Without Accountability
For task designers situated in industry, the structure of professional learning around risk disclosure was even more informal. Rather than compliance systems or ethical audits, task designers often learned disclosure practices through peer modeling and internal documentation. P21 explained that they were given task templates that included risk disclosure from a supervisor, which they reused without modification. P13 described how their company had internal guidelines, but added that \textit{``there would be no disciplinary action''} if someone actually failed to follow them. These accounts reflect a culture of local, non-binding norms, where designers took cues from mentors or colleagues but were rarely held accountable by the organization itself. Thus, even in companies that promoted ethical values, the implementation of disclosure was uneven.  Without centralized oversight, risk disclosure practices were shaped by what was modeled, not mandated. This created wide variability in how seriously risk disclosure was treated by task designers, potentially further fragmenting the landscape of responsible design practices across the AI workforce.
Overall, professional norms in academia and industry helped raise awareness of risk but did little to support actionable or consistent disclosure. Learning was often informal and observational, making ethical practice contingent on what was modeled rather than mandated.

\subsection{Risk Disclosure as a (Perceived) Tradeoff? Balancing Risk Disclosure and Data Integrity}%{Tensions Between Risk Disclosure and Data Quality}

% \begin{table}[h]
% \centering
% \begin{tabular}{|p{4.5cm}|p{9cm}|}
% \hline
% \textbf{Quality Concern} & \textbf{Relation to Risk Disclosure} \\
% \hline
% Task Designer Reputation & Transparent communication about risks can improve worker trust and willingness to participate in future tasks. \\
% \hline
% Data Integrity & Disclosing sensitive content too explicitly may bias responses or reduce the validity of experimental outcomes. \\
% \hline
% Worker Preparedness & Providing upfront warnings may enhance worker resilience and data completeness, though designers must balance clarity with the risk of scaring workers off. \\
% \hline
% \end{tabular}
% \caption{Quality-related considerations influencing risk disclosure.}
% \label{tab:quality_considerations}
% \end{table}

While contextual factors like organizational norms and platform affordances shaped task designers’ disclosure strategies, concerns about data quality, particularly in relation to training or evaluating AI systems, emerged as a dominant force. Designers often weighed risk disclosure against the perceived need for precise, representative, and bias-free data, revealing tensions between protecting worker well-being and optimizing for system performance.

\subsubsection{Risk Disclosure as Instrumental to Data Quality}
For many task designers, maintaining a positive reputation among crowd workers was not only framed as an ethical gesture but also as a practical necessity for recruiting attentive, qualified contributors. Reputation reportedly had direct implications for data quality, which was especially important in RAI contexts where subtle errors could degrade downstream model performance \cite{zhang2025making}.
% Risk disclosure was often framed as a way to build credibility with workers and ensure faster, higher-quality responses, turning ethical transparency into a practical investment in long-term task success. 
Importantly, task designers did not describe this as a form of algorithmic management (i.e., ~\cite{jarrahi_platformic_2020, lee_working_2015}), as they did not express concerns about being rated poorly or reported to the platform by an unhappy crowdworker. Rather, efforts to build a positive reputation were described as emerging through informal, interpersonal channels, such as informal social networks \cite{Wood_2019Disembeddedness_of_labor} or off-platform sites like Turkopticon \cite{irani2013turkopticon} or Reddit, rather than through structured feedback systems built into the platform. 

A key motivating factor for task designers to disclose risk was to build a positive reputation on crowdsourcing platforms, which could, in turn, make their tasks appealing to qualified workers and help them obtain high-quality work. In these accounts, risk disclosure was not only an ethical practice but also a strategic decision to foster worker trust and engagement, which was thought to improve the reliability and richness of responses.
P19 described how investment in ethical practices like content warnings and support services was ultimately beneficial to data quality and study success: \textit{``while providing warning and supports...may require some upfront investments, it can also lead to like long term benefits, such as...improved data quality...increased worker engagement and enhance[d] reputation, while also ensuring compliance and ethical guidelines.''} For this participant, reputation was not a static social credential, but something that actively shaped how well a task would perform in practice.
Others emphasized that reputation among workers on platforms like Mechanical Turk could influence how quickly and attentively tasks were completed. P1 explained: \textit{``Turkers can see the people who...posted the assignments...some people have a good reputation among the Turkers for...paying fairly and accepting...most work...[so] their work gets picked up fairly quickly.''} Here, a good task designer reputation was tied not just to speed, but implicitly to worker motivation and effort.
In these cases, task designers framed their risk disclosing behavior as part of a feedback loop: ethical and transparent practices contributed to a stronger reputation, which led to better worker engagement, and ultimately, higher quality data.

%  new worker preparation section
Disclosure was also seen as a tool for helping workers feel prepared to engage with difficult content, though designers debated how much to reveal without compromising either data quality or worker well-being.
Several task designers sought to balance worker preparation with the need for clean, untainted data, especially in tasks involving model auditing, safety evaluation, or LLM fine-tuning. While some believed that preparing workers led to higher-quality labels, others worried that disclosure would deter participation or compromise task realism, both of which could affect model performance benchmarks.
% Crucially, disclosure was also seen as a way to directly or indirectly ensure higher---quality data by helping workers feel prepared for the task. 
P1, who believed they provided detailed information about the type of content in their tasks, noted that \textit{``all of these things contribute to getting high quality human labels.''} P13 reasoned that unprepared workers might disengage or struggle to provide useful input: 
\begin{quote}
    \textit{``[If] the question is traumatizing because you have to remember what happened in the past that you're trying to forget and you're told to explain [this experience] in detail, you [may] just [type a single] word...[not] being detailed with your explanation and all that means maybe you're trying not to remember your past [experiences].''}
\end{quote}
However, task designers also described tradeoffs between preparing workers and preserving the integrity of their study designs. Though Prolific allowed task designers to filter by workers who had opted into sensitive tasks, P5 chose not to use these filters because \textit{``[they] needed [workers] to be sensitive to a particular extent. So [they] felt like when [they] turn off those sensitivity [filters] they don't get that raw information.''}

Others grappled with this tension even more cautiously. P3 framed it as a balance between ethical responsibility and methodological soundness:
\begin{quote}
    \textit{``The ethicist in me says, if it's toxic content that can harm people or have impact on their mental health, definitely err on the side of figuring out a way to provide some content warning even if it's just very abstract...[but] because there's a lot of experimental design that could potentially be biased by giving that warning, for certain studies, I might pick that design as an option of last resort.''}
\end{quote}
As P8 echoed, abstract or vague warnings could serve as a middle ground: \textit{``You don't say up front in the description [that] you are collecting concerns [about sensitive topics]...you want to say you are collecting options about a computer system,''} while still acknowledging, \textit{``there's some tensions over there [as] you need to protect your study task designers to a certain extent for doing this study.''}
Some task designers also tried to interpret signals of worker preparedness or dropout behavior based on task completion patterns. P11 described how they inferred worker disengagement when workers failed to respond to reminders or abandoned the task altogether: \textit{``Some of them, I will just assume maybe they are not available, or they are not...interested anymore in the task.''} While not explicitly framed as a response to risk, these observations suggest that some designers used patterns of non-completion as a signal that interaction with sensitive content could have led to some type of negative reaction from workers, deterring them from completing the task.

Taken together, these accounts illustrate how worker preparation was seen as integral to data quality, but also deeply entangled with methodological tensions. Task designers weighed whether and how to disclose risk based on assumptions about what would help workers stay engaged without compromising the validity of their responses. These quality-based decisions often eclipsed formal ethical guidance, suggesting that in AI development, the imperatives of performance and scalability may take precedence over consistent risk communication. As AI systems increasingly rely on crowd labor to simulate edge cases or evaluate sociotechnical impact, understanding how designers navigate these trade-offs is critical to shaping more accountable data pipelines.

Task designers described disclosure not only as an ethical practice but also as a strategic one that was key to building worker trust, enhancing participation, and improving label quality. This reframes disclosure as instrumental to data integrity, rather than a threat to it. What's more from this viewpoint, well-prepared workers were perceived to produce richer, more reliable data, but designers struggled to balance preparation with methodological rigor. These tensions illustrate how disclosure could be both a support and a potential liability for task performance.

\subsubsection{Risk Disclosure as Harmful to Data Integrity}
Task designers developing training data for AI systems often expressed concern that disclosing too much about task content would introduce priming effects or sampling bias. These concerns reflect the logic of ``performative accuracy'' in AI development~\cite{zhang2025making}, where the perceived objectivity of model performance is tied to the controlled conditions under which data is collected.
For example, P3, an academic who designed tasks asking workers to annotate bias to assist with the development of AI tools that prevent job posting biases, emphasized the tension between ethical transparency and methodological rigor: \textit{``what I'd recommend is...for a researcher to think about what information might affect the responses that you're getting...and how that could bias your actual study results, which would be not great science.''} They noted the importance of making workers aware of the subject matter in general terms so they could make informed decisions, but cautioned against revealing details that might shape responses: \textit{``the tradeoff is...disclosing as much about the subject matter...that would be important for the worker to know...[but not] bias your actual study results.''}
Similarly, P8 worried about priming effects, particularly in studies where unfiltered or spontaneous reactions were critical: \textit{``you may then prime people right in advance...to think of certain things in a certain way...so we hedged that often...with things like, `your opinion about potentially...' using some of these kinds of hedging language.''} In such cases, task designers sought to balance disclosure with careful framing to reduce interpretive bias.

P6 also reflected on the emotional risks of early disclosure, suggesting that revealing too much upfront might overwhelm or disorient task designers, especially in initial encounters with sensitive content: \textit{``It could be discrimination. [Task designers] might receive...responses that might not really be appealing for a start. People tend to be emotionally affected by responses...when they're just starting...So I feel that it should just be left first, then probably the second or third time [disclosure is introduced].''} Here, withholding information was framed not only as a design strategy but as a form of emotional pacing to avoid response distortion.

Across these perspectives, task designers viewed risk disclosure as closely tied to the design goals and epistemic commitments of their studies. While some feared that content warnings would bias or suppress authentic responses, others saw disclosure as necessary to prevent emotional shutdowns, confusion, or abandonment that could equally threaten data quality. These competing views reflect task designers’ uncertainty about what constitutes `quality' data — an uncertainty intensified by a field-wide focus on data quality in AI development \cite{zhang2025making}, often at the expense of other considerations, such as ethics, which are sometimes overlooked or even positioned in opposition to data quality.

\section{Discussion}

\subsection{Who Cares for Crowdworkers? Re-locating Accountability for RAI Crowdworker Well-being}

The critical work of making AI ``responsible'' requires an army of crowdworkers to annotate potentially undesired content~\cite{zhang2024aura, wang2022whose}, which may contain hate speech, violence, child sexual abuse materials, or other kinds of sensitive content and AI behaviors. When such content is shown to crowdworkers, it may severely and negatively impact their well-being. Disclosing the risks involved in sensitive content work tasks can help reduce this negative impact \cite{bharucha2023content}. Decisions to disclose or not disclose risks are a consequential yet understudied part of the much bigger invisible human infrastructure shaping the behavior of AI systems~\cite{gray2019ghost, zhang2024aura}.

Our study illustrates the range of task designers' risk disclosure practices and attitudes regarding sensitive tasks that unfold across the RAI task lifecycle. These practices are shaped by individual beliefs about their duty of care toward crowdworkers’ well-being, a hodgepodge of platform safeguards and organizational guidelines, and tensions regarding data quality. Therefore, our work emphasizes the task designer-worker interface as a critical site of situated decision making around risk disclosure.

\subsubsection{\edit{Displaced Responsibility and the Burden on Task Designers}} 

%Task lifecycle:
Our findings show that some task designers consider risk from the early stages of task conceptualization, while others only do so when prompted by external factors such as their organization or the crowdsourcing platform. This variation underscores the lack of sustained support for risk awareness throughout the RAI task lifecycle.
This discontinuity in support implies that current organizational and platform mechanisms position risk disclosure as a procedural checkpoint rather than a continuous ethical responsibility. 
% The fact that some task designers only recognized harm through downstream signals, like dropout rates or worker complaints, also underlines the importance of designing systems and organizational accountability mechanisms that support ethical disclosure along the entire task lifecycle, especially that support crowdworker feedback and agency \cite{gegenhuber_microphones_2021}, instead of only at the conceptualization of the task, as many ethics and compliance review processes do.
The fact that some task designers only recognized harm through downstream signals, like dropout rates or worker complaints, also underlines the importance of designing systems and organizational accountability mechanisms. These mechanisms should support ethical disclosure and enable crowdworker feedback and agency \cite{gegenhuber_microphones_2021} along the entire task lifecycle, instead of only at the conceptualization of the task, as many ethics and compliance review processes do.
Treating disclosure as a dynamic design process would help identify the points at which task decisions need better support to mitigate risk for workers.

%Individual values:
Highlights from our analysis, such as that of task designers invoking their own values around transparency and ethics to guide decisions to disclose risk, reveal how the personal responsibility task designers feel plays a large role in how they make decisions to disclose risk. 
The reliance we observed on individualized ethics reflects what Ali et al. describe as ``ethics of entrepreneurship''~\cite{ali_walking_2023}, where designers may assume moral accountability in the absence of external requirements. However, this sense of duty was rarely complemented by structural safeguards, which oftentimes left task designers to navigate task disclosure decisions alone. As such, we extend critiques of flaws in ethics or institutional review processes (i.e., IRB review)~\cite{Cascio03042018, zook_ten_2017}, while also drawing parallels from prior work that shows how mid-level actors within organizations face certain constraints that limit their agency~\cite{ali_walking_2023}.

%Organizational dynamics: 
Moreover, our findings illustrate that organizational structures may offer little sustained support for task designers to manage risk disclosure, leaving ethical decisions to be improvised rather than institutionalized.
Building on concerns raised Vitak et al.\cite{vitak_beyond_2016}, these insights highlight the need to complement the personal ethical responsibility that designers felt with structural safeguards that offer continual support, guidance, and standardization. However, as cautioned by prior research \cite{turri2024transparency}, it is possible that even well-intentioned transparency frameworks and guidelines~\cite{diaz2022crowdworksheets,gebru2021datasheets, holland2020dataset} may fall short when deployed within fragmented or absent organizational infrastructures. This raises critical questions about how effectively such frameworks can be implemented in practice without more consistent institutional backing.
In the academic context specifically, our findings support growing calls to reconsider how Institutional Review Boards (IRBs) evaluate research involving crowdworkers \cite{kaushik2023resolvinghumansubjectsstatus}. We urge that, particularly for sensitive RAI tasks, IRBs give greater consideration to recognizing crowdworkers as human subjects, in light of the distinct and often underacknowledged harms such tasks may pose to their well-being.

Importantly, our findings also point to the mobility of task designers across professional contexts, from academia to industry and vice versa, as a possible vector for change. While structural accountability for crowdworker well-being remains uneven, individual designers often carry norms and habits from one setting into another~\cite{pedersen2024cultivating}. In this sense, academic environments, despite their own gaps, can serve as early sites for cultivating disclosure awareness. Embedding stronger standards for ethical risk mitigation in research training or review practices may thus have cascading effects, as designers carry these values into industry roles where formal accountability structures are even weaker.

%Platform: 
Our findings also suggest that platform features intended to support risk disclosure are inconsistently used, in part due to the absence of standardized enforcement or shared norms. This variability was evident not only in how task designers selectively engaged with available tools, but also in how they repurposed unrelated platform elements, a practice Duguay describes as ``off-label'' use \cite{duguay_you_2020}, such as leveraging task keyword fields to communicate risk. These adaptations reveal that even well-intentioned affordances may fail to ensure ethical practice when designers are left to interpret their use without guidance.
This absence of normative and structural support is further compounded by platforms’ frequent commitments to maintaining a stance of neutrality \cite{Gillespie_politics}, which may insulate them from responsibility when harm \edit{is inflicted. Platforms tend to shrug responsibility rather than to share it \cite{Helberger01012018}. }In line with Widder’s concept of dislocated accountability \cite{widder_dislocated_2023}, our findings illustrate how responsibility for crowdworker well-being becomes further fragmented across individual designers, \edit{platforms}, \edit{institutions, and governments}. \edit{Our emphasis on task designers in this analysis should therefore be read as an effort to make visible how responsibility for worker well-being risk is displaced onto individual task designers in the absence of robust organizational and platform governance.}

\subsubsection{Data Quality vs. Disclosure: A False Binary \edit{for RAI Content Work}?} 
%Data quality: 
The relationship between disclosure and data quality was perhaps the most tense and varied of all. Faced with organizational pressures to efficiently deliver high-quality data \cite{Sambasivan2021}, made even more prescient with the extremely high levels of data accuracy purportedly necessary for AI development~\cite{zhang2025making}, academic and industry task designers alike prioritized data quality over almost everything else. 
\edit{For some, greater transparency to workers was seen as compatible with this goal, increasing interest in the task and fostering trust between workers and task designers. For others, our findings suggest the inverse: they feared that disclosure would bias their sample and reduce the quality of data produced. Such fears may be related to prior work showing that factors such as worker demographics can influence data annotations~\cite{ding2022impact}. We propose that, in the context of RAI content work, where workers were hired to label violent content, moderate disturbing text, or simulate harmful behavior, this decision is a false binary that has festered in the vacuum of structural ethics guidance. Primarily, there is a question of how we define data quality and whether that should be constrained to narrower metrics, such as agreement in data labels, or more broadly encompass factors indicating the retention and well-being of workers, as highlighted in recent CSCW literature~\cite{zhang2024aura}. Moreover, there is limited research demonstrating the negative effects of risk disclosure on such factors, with prior work primarily focusing on examining the effects of differing worker demographics rather than worker well-being, for example~\cite{kashima2024trustworthy, wan2023everyone}. As such, overlooking the importance of risk disclosure for RAI content work may further contribute to the continued professional diminishment of the practice \cite{Sambasivan2021}.
This attitude reinforces a narrow notion of objectivity in AI development, in which ethical considerations are treated as a contaminant rather than a necessary condition for valid knowledge production. However, we note that outside the specific context of risk disclosure for crowdsourced RAI content work, there exist instances (e.g., experiments involving deception) where a thorough post-hoc debriefing for participants might be more appropriate than upfront risk disclosure. } 

Our findings reveal a tension at the heart of responsible AI labor: while task designers are often acknowledged as the ethical gatekeepers of AI data quality, they are rarely equipped with the infrastructure, guidance, or institutional support required to sustain that responsiblity. As we show, task designers across industry and academia struggled without standardized guidance, often relying on intuition or norms. Their decisions around risk disclosure were shaped in a vacuum through which platforms claim neutrality and institutions defer oversight, ultimately diffusing accountability. 

We emphasize the sheer disproportionate responsibility task designers hold in protecting worker well-being within the RAI development pipeline. The task-designer-worker relationship in our study emerged as an under-examined yet critical node where risk could be actively mitigated through situated design choice. Unlike other actors in the RAI development environment, task designers are able to have direct contact with terms of the worker experience. In fact, it is this concentration of responsibility that offers a unique opportunity. In a development supply chain that is often marked by distributed accountability~\cite{widder_dislocated_2023}, task designers represent an important pivot point for risk disclosure. %While we do not argue that task designers should be the sole bearers of this responsibility, we locate them as the most proximate actors to crowdworker risk. 
Recognizing this reality offers us a pragmatic intervention point which we may later use as an opportunity for targeted support and tooling that could alleviate the burden task designers face while improving work experiences for workers. 

However, by viewing risks to crowdworker well-being from the situated knowledge of task designers, rather than from the "panoptic" perspective of top-down transparency initiatives~\cite{widder_dislocated_2023}, we lay the groundwork for future research that takes advantage of the power that internal stakeholders possess. In lieu of more ambitious structural changes to distribute accountability across platforms, organizations, and individuals, design interventions could leverage this rare instance of ethical agency within the AI supply chain.

Still, interventions alone are not enough. As Lilly Irani notes in her retrospective on Turkopticon~\cite{irani2016stories}, mutual aid and increased transparency among workers did little to curb platform exploitation when not accompanied by structural change. In our study, we observed a parallel dynamic: task designers raised concerns about the ethics of their work, but their corrective actions remained constrained by institutional inertia, platform neutrality, and the relentless pace of development. Without broader shifts in how responsibility is distributed and enforced, \edit{like those suggested by Helberger et al. \cite{Helberger01012018},} design interventions—no matter how well intended—risk reinforcing the very dislocation they seek to remedy.

\subsection{Risk Disclosure as a Design Space} 
Our findings suggest that risk disclosure is not simply a matter of cultivating ethical clarity or increasing compliance: it is a design problem. Decisions about whether and how to warn crowdworkers of well-being risk are embedded in the micro-structures of task creation (i.e., in how prompts are framed, how task interfaces are built, etc.). In this light, disclosure is not a singular act but an iterative process, and is shaped by tools, workflows, platform governance, and affordances~\cite{diaz2022crowdworksheets, holland2020dataset, pushkarna2022data, toxtli2021quantifying, rzeszotarski2012crowdscape}. Yet RAI task design environments \edit{provide limited and often inconsistent support for these decisions.}

% empirical evidence from findings + comparative analysis of platform affordances for why it's a design problem
Many task designers in our study relied on improvised strategies to identify, frame, and communicate risk. In the absence of clear design cues or supportive workflows, they turned to personal intuition, peer templates, or ad hoc revisions made during task deployment. Notably, this improvisation persisted despite the availability of platform affordances \edit{on widely used crowdsourcing sites such as Prolific} to present task information, review worker outputs, and sometimes collect feedback \cite{rzeszotarski2012crowdscape, gegenhuber_microphones_2021, duguay_you_2020}, as well as community-built resources such as worksheets and requester templates \cite{diaz2022crowdworksheets, bragg2018sprout}. \edit{Across platforms, tools vary: Prolific provides comparatively structured fields for flagging explicit or sensitive content and for screening and consenting workers, while controls remain more limited on sites such as MTurk \cite{prolific2025sensitive, prolific2025participant, mturk2018aup}. Yet even on Prolific, some participants chose to bypass pre-screening options out of concern that they might exclude relevant workers, revealing that existing features do not fully align with task designers’ needs for risk disclosure.} External channels like Turkopticon further illustrate how feedback often travels outside the platform rather than through it \cite{irani2013turkopticon}. \edit{Taken together, these examples reveal a kind of ``design tinkering'' in which task designers improvise around uneven and sometimes underutilized platform tools, struggling to disclose risk meaningfully while still ensuring they can gather quality data efficiently~\cite{papoutsaki2015crowdsourcing, gutheim2012fantasktic, kittur2008crowdsourcing}. While existing tools represent a promising start, the gap between task designer needs and innovation in tools available points to a deeper design need: disclosure practices must be integrated more thoughtfully into task workflows in ways that are visible, actionable, and aligned with the practical realities of crowdwork. Rather than leaving task designers to navigate these tradeoffs alone, the space of well-being risk disclosure presents a concrete design opportunity. Tools might include structured prompts, customizable warning templates, or lightweight pre-review mechanisms. Importantly, such tools must be flexible enough to accommodate the diversity of crowdwork contexts, yet robust enough to encourage designers to adopt more consistent and thoughtful disclosure practices~\cite{bragg2018sprout, diaz2022crowdworksheets}.}

% Our findings show that even when platforms offer structured tools for risk disclosure, task designers are often unaware of these features, uncertain how to use them, or unconvinced of their utility. While these tools bridge the gap between feature availability and designer uptake, they point to a deeper design need: disclosure practices must be integrated more thoughtfully into task workflows in ways that are visible, actionable, and aligned with the practical realities of crowdwork. Rather than leaving task designers to navigate these tradeoffs alone, the space of well-being risk disclosure presents a concrete design opportunity. There is a real need to build tools that help task designers calibrate how much to disclose, when, and through what means, depending on task goals and specific types of sensitive content involved in the task. These might include structured prompts, customizable warning templates, or lightweight pre-review mechanisms. Importantly, such tools must be flexible enough to accommodate the diversity of crowdwork contexts, yet robust enough to push designers toward more consistent and thoughtful disclosure practices~\cite{bragg2018sprout, diaz2022crowdworksheets}.

% designing for better disclosure doesn't have to compromise data quality
Critically, positioning disclosure as a design problem also clarifies its relationship to data quality. As task designers in our study noted, disclosure is often weighed against concerns about sample bias or worker priming. But well-designed, timely communication of risk may actually improve data quality by reducing attrition and adverse selection from opaque rejections~\cite{mcinnis2016taking}, supporting more accurate worker understanding, and encouraging stable task engagement. Just as ambiguity in task instructions can introduce noise, ambiguity around emotional risk can contribute to dropout or distorted responses~\cite{bharucha2023content}. Designing risk disclosure directly into task workflows can thus benefit both ethical outcomes and dataset integrity. 

% accounting for governance challenges in design
 \edit{Our solution is pragmatic: it is meant to improve conditions for workers in lieu of stronger platform governance measures. This is not to say that better design replaces the need for stricter governance, but rather it offers a complementary approach.} \edit{By embedding disclosure into existing checkpoints and tying it to familiar metrics such as completion rates and worker retention, these tools can be framed not only as ethical safeguards but also as strategies for reducing costly attrition and rework. At the same time, we caution that design interventions alone, particularly those focused on individual task designers, must not distract from broader structural responsibilities. Interventions at the task designer level cannot fully address systemic power imbalances, opacity in platform operations, or the absence of enforceable labor protections. Instead, they are most effective when paired with governance mechanisms and institutional incentives that prioritize worker well-being.} Ultimately, treating disclosure as a design challenge invites new possibilities for workflow innovation, prototyping, and ethical reflection, not as a one-time obligation, but as a situated, ongoing process embedded in the work of building responsible AI systems.

\subsection{Limitations and Future Work}
While this study offers an in-depth empirical account of how task designers navigate risk disclosure in crowdsourced AI annotation, some limitations are inherent to the methodological choices we made. While we acknowledge that these choices were necessary to surface situated, interpretive practices that are otherwise difficult to observe, we also provide additional clarification to highlight opportunities for future work. 

First, one potential concern is selection bias: task designers who are already ethically attuned or sensitive to worker harm may have been more willing to participate in a study focused on well-being. To mitigate this, our recruitment messages framed the study as a practical opportunity to improve the RAI task design process for both designers and workers, rather than emphasizing harm, ethics or risk reduction. This neutral framing helped avoid recruiting only individuals with a strong preexisting commitment to responsible AI or worker protection. \edit{However, we acknowledge limitations still present in our approach, such as the phenomenon of social desirability bias~\cite{bergen2020everything}.} \edit{Moreover, our sample reflected a diversity of ethical orientations: while some participants demonstrated a high awareness of potential harms, others (e.g., P20 and P21) described limited engagement with risk considerations or adherence to top-down directives without engaging in} personal ethical reflection. This range suggests that our participant pool included \edit{individuals with varied levels of reflexivity, addressing concerns of sampling bias} toward highly risk-aware individuals. \edit{Regardless, future work may further explore phenomena such as widespread challenges task designers face through alternative methods such as quantitative surveys.} 

Second, our participant pool, although diverse in terms of role and platform experience, primarily consisted of designers working within academic, research, and small organizational contexts. Additionally, our sample primarily consisted of participants from the United States. As such, our findings may not fully capture the workflows and constraints experienced by task designers embedded in large-scale commercial annotation operations, especially those shaped by different institutional logics or regional labor frameworks. Extending this work to include task designers situated within global enterprise-level AI pipelines could reveal how risk is negotiated under tighter production pressures and across geographic, linguistic, and cultural contexts, particularly in regions where crowdsourcing platforms or AI norms differ significantly.

Third, like many interview-based studies~\cite{yin2015qualitative}, we rely on self-reported practices to access participants’ reflections, reasoning, and decision-making processes, elements that are rarely visible through artifacts alone. This approach was essential for eliciting designers’ reflections, intentions, and ethical reasoning---elements that may not be directly observable through task artifacts alone. At the same time, interview data reflect how participants recall and frame their practices, which may differ from their real-time decision-making. Given the sensitivity of topics such as risk, ethics, and responsibility, some participants may have unintentionally minimized or downplayed omissions or emphasized their best intentions. Future work could complement our findings with mixed methods approaches, such as observational studies, design artifact analysis, or interface walkthroughs, to triangulate reported accounts with actual task creation behavior. As a next step, future work could take the form of participatory design or prototyping studies that engage task designers in developing, testing, and refining features that support meaningful risk disclosure. Doing so would help move disclosure from a discretionary practice to a more intentional, structured, and accountable part of RAI data pipeline design. \edit{Additionally, further research may explore downstream impacts of risk disclosure on worker well-being.}

\section{Conclusion}
This paper began with a question about responsibility: who recognizes, discloses, and mitigates the risks faced by the human labor behind AI systems? Although prior work has rightly centered workers themselves, we shift analytical attention to a rarely examined site \edit{with the potential for ethical action}: the task designer.
Through interviews with task designers across academic and industry settings, we found that risk disclosure is rarely supported by policy or platform. Instead, task designers navigate a landscape shaped by institutional neglect, conflicting pressures, and personal ethics. They tinker, intuit, and adapt in a way that makes risk disclosure not a rule, but a design improvisation under constraint.
In surfacing these practices, we contribute to ongoing efforts in CSCW and the broader HCI community to study invisible labor, platform governance, and human-in-the-loop AI efforts. We argue that meaningful risk disclosure is not merely a question of transparency or compliance, but a situated design challenge--one that demands better tools, norms, and institutional support.
As the burden of AI responsibility continues to fall on the crowd, we must ask what it would take to support task designers not only in building accurate systems but in caring for the workers who power them. Locating accountability in task design means not just expecting more from individuals but reimagining the infrastructures that shape their decisions.

\section{Acknowledgments}
This project was supported through Microsoft's AI and Society Fellowship\footnote{https://www.microsoft.com/en-us/research/academic-program/ai-society-fellows/}. Our work specifically focused on supporting the responsible AI red teaming human infrastructure. The first author was additionally supported by the NSF GRFP DGE2140739. We note that we did not recruit from sponsoring organizations, and that our views are our own and do not represent those of our supporters or sponsors. We also thank Mary L. Gray for always advocating for our work and support in the early stages of the project. Finally, our work would not have been possible without the time and thoughtful engagement of our participants.

\bibliographystyle{ACM-Reference-Format}
\bibliography{_references, _references_aura, _references_crowdsourcing, _references_worker_discretion}
\section{Appendix}

\subsection{Study Recruitment Questionaire}
We include questions participants answered in a questionaire they completed to indicate interest in participating our study. Note that participants had the option to not answer any of these questions.

\begin{enumerate}
    \item \textbf{Email address} \\
    \underline{\hspace{10cm}}

    \item \textbf{What country are you currently located at?}
    \begin{itemize}
        \item United States
        \item Other
        \item Prefer not to answer
    \end{itemize}

    \item \textbf{Please provide your job title. If you prefer not to answer, enter ``NA.''} \\
    \underline{\hspace{10cm}}

    \item \textbf{What type of organization do you work in? (we will only ask for a broad category for privacy purposes)}
    \begin{itemize}
        \item Academia (e.g., university)
        \item Industry (e.g., for-profit company)
        \item Non-profit
        \item Other
    \end{itemize}

    \item \textbf{If you selected ``other'' in the task above, please fill in the blank.} \\
    \underline{\hspace{10cm}}

    \item \textbf{What is the approximate size of your organization (how many employees)?}
    \begin{itemize}
        \item 25,000 and more
        \item 5,000 -- 24,999
        \item 1,000 -- 4,999
        \item 250 -- 999
        \item 50 -- 249
        \item 10 -- 49
        \item Prefer not to answer
    \end{itemize}

    \item \textbf{What type of tasks do you have experience requesting/will you request?}
    \begin{itemize}
        \item Data generation (e.g., provide examples of toxic language)
        \item Data labeling (e.g., which of these is an example of sexual content)
        \item Content removal (e.g., is this content harmful enough that a model should not produce it?)
        \item Adversarial prompting (e.g., provide a prompt for a model to produce violent content)
        \item Prefer not to answer
        \item Other
    \end{itemize}

    \item \textbf{How many responses do you typically aim for in your tasks?}
    \begin{itemize}
        \item 100,000+
        \item less than 100,000
        \item less than 10,000
        \item less than 1,000
        \item less than 100
        \item Prefer not to answer
    \end{itemize}

    \item \textbf{What types of potentially sensitive content have been in tasks you requested? Please select all that apply.}
    \begin{itemize}
        \item Misinformation
        \item Bias/stereotyping
        \item Hate speech
        \item Violent and graphic content
        \item Harassment and Bullying
        \item Terrorism and Extremism
        \item Sexually explicit content
        \item Self-harm and suicide
        \item Illegal activities
        \item Scams and fraud
        \item Child endangerment (CSAM content)
        \item Other
        \item Prefer not to answer
        \item None of these apply to tasks I have requested
    \end{itemize}

    \item \textbf{If you selected ``other'' in the task above, please fill in the blank.} \\
    \underline{\hspace{10cm}}

    \item \textbf{What crowd source platforms have you utilized/do you plan to utilize to request tasks?}
    \begin{itemize}
        \item Amazon Mechanical Turk
        \item Prolific
        \item Clickworker
        \item Remotasks
        \item DataAnnotation
        \item TaskUs
        \item Other
        \item Prefer not to answer
    \end{itemize}

    \item \textbf{If you selected ``other'' in the task above, please fill in the blank.} \\
    \underline{\hspace{10cm}}

    \item \textbf{Please provide an example of a task you requested (e.g., asking people to annotate sentences for gender discrimination).} \\
    \underline{\hspace{10cm}}

    \item \textbf{Upload a screenshot of the task you requested if possible.} \\
    \textit{(Include as attachment if digital, or describe here if printed)} \\
    \vspace{1cm}
\end{enumerate}

\subsection{Semi-Structured Interview Protocol}
In the following section, we detail the semi-structured interview protocol we used. 
\subsubsection*{Background and Context}
First, we will be asking you some questions about your experience going into responsible AI crowd work. As always, all questions are optional and feel free to skip them if you are not comfortable answering.

\begin{enumerate}[label=\arabic*.]
    \item Please introduce yourself. What is your first/preferred name, and current occupation?
    \item Please define Responsible AI in your own terms.
    \item How did you become involved in data enrichment or Responsible AI work?
    \item What role does it play in your job?
\end{enumerate}

\subsubsection*{Task Setup and Motivations}
We are very curious about the process requesters go through in setting up crowd work tasks. If you are able to do so, please recall a recent experience you have creating a task. Assume we do not have any knowledge about requesting tasks.

\textit{Potential follow-up questions:}
\begin{itemize}
    \item What is the task for? (e.g., a model, RLHF)
    \item What is the overall goal you have where the crowdwork fits in?
    \item What platform(s) did you choose to request these tasks?
    \item What informed your decision to choose the platform(s) to request these tasks?
    \item What other procedures do you engage with before posting the task (e.g., testing task with workers, approval with management)?
\end{itemize}

\subsubsection*{Task Information}
Thank you for sharing that, that was very insightful. We would like to learn more about the details of the task with some follow-up questions. There are absolutely no right or wrong answers. We are simply trying to explore the space for collaboration in helping requesters like yourself better design tasks. Feel free to skip any questions.

\textit{Follow-up questions:}
\begin{itemize}
    \item What are the main components of information you include?
    \item What do you include in the pre-task description versus the actual task?
    \item How much of the information included is your decision (versus an organizational requirement)?
    \item Do you work with others to write the task?
    \item Are you able to reveal how much you typically compensate workers (skip if not)?
    \item What do you think is most important to tell crowd workers about your tasks?
    \item Do you have any existing protocols, frameworks, or tools to guide this process?
    \item What risks do you see your task posing to workers?
    \item Are there any requirements from your organization to disclose risks to workers? If so, can you share your experience with this?
\end{itemize}

\subsubsection*{Task Management Process}
Now, we would like to know more about what happens after a task is posted.

\begin{itemize}
    \item Once a task has been posted, what do you do to monitor it?
    \item Have you ever edited tasks to include more or better information?
    \item What communication channels did your chosen platform(s) offer for contact with workers?
    \item How did this influence your experience with the task? (e.g., better quality data, ability to recruit repeat participants, or added difficulty)
    \item How do you decide on the amount you paid workers?
    \item Have you ever had to change this number and why?
\end{itemize}

\subsubsection*{Task Debriefing}
At times, we know that debriefing after a task is necessary. We have a few follow-up questions about this.

\begin{enumerate}[label=\arabic*.]
    \item Can you start by describing to us what you tell workers after they complete a task? It can be as simple as saying thank you and sending them on their way to get paid.
\end{enumerate}

\textit{Potential follow-up questions:}
\begin{itemize}
    \item What information did you include for workers to view after they completed the task?
    \item What did you do with the task data/output?
    \item Do you think it is important to tell workers what is done with their data/work?
    \item Do you think it is important for workers to be given opportunities to provide feedback on tasks?
    \item Do you think it is important for workers to have opportunities to refute decisions made (e.g., rejections)?
\end{itemize}

\subsubsection*{Challenges}
Lastly, we would like to take a moment to reflect on challenges you face as a requester in this space. Again, there are absolutely no right answers we are seeking. We will start with some general questions:

\begin{itemize}
    \item What challenges do you face in providing transparency about your tasks?
    \item What are the challenges in each phase: pre-task screening, during the task, and post-task?
    \item What aspects of the task creation process would you change to ensure greater transparency?
    \item Who can help you make this change (e.g., what can platforms do)?
    \item If you could have the perfect environment (wave a magic wand) for requesting tasks, what would it be?
    \item Who would be there supporting you (e.g., managers)?
    \item What resources would you have (e.g., more money to pay workers)?
\end{itemize}

\end{document}